\documentclass[conference]{IEEEtran}

\usepackage{amsmath,amssymb,amsthm}
\usepackage{graphicx}
\usepackage{cite}
\usepackage{booktabs}
\usepackage{multirow}
\usepackage{array}
\usepackage{color}
\usepackage{url}
\usepackage{tikz}
\usepackage{fancyhdr}

\usepackage{subcaption}

\usetikzlibrary{arrows.meta, shapes, positioning}

\begin{document}

\title{DRASTIC: A Dynamic Resource Allocation Framework over 6G Network Slicing in Task-aware Closed-Loop Tactile Internet Applications}

\author{\IEEEauthorblockN{
\large Narges Golmohammadi, Madan Mohan Rayguru and Sabur Baidya\\
\normalsize Department of Computer Science and Engineering, University of Louisville, KY, USA}
\normalsize {e-mail:  narges.golmohammadi@louisville.edu, madanmohan.rayguru@louisville.edu, sabur.baidya@louisville.edu}
\vspace{-6mm}
}

\maketitle
\thispagestyle{fancy}   
\pagestyle{fancy} 

\fancyhf{}
\pagestyle{fancy}
\thispagestyle{fancy}
\renewcommand{\headrulewidth}{0pt}
\renewcommand{\footrulewidth}{0pt}

\setlength{\headheight}{14pt}
\setlength{\headsep}{10pt}

\fancyhead[C]{%
\footnotesize\makebox[\textwidth][c]{This article has been accepted for publication in IEEE DySPAN 2026.}}

\fancyfoot[C]{
    \begin{tikzpicture}[remember picture, overlay]
        \node[anchor=south,yshift=10pt] at (current page.south) {
            \parbox{\textwidth}{
                \centering
                \footnotesize
                \textcopyright~2026 IEEE. Personal use of this material is permitted. Permission from IEEE must be obtained for all other uses, in any current or future media, including reprinting/republishing this material for advertising or promotional purposes, creating new collective works, for resale or redistribution to servers or lists, or reuse of any copyrighted component of this work in other works.
            }
        };
    \end{tikzpicture}
}

\begin{abstract}
This work proposes a novel learning driven bandwidth optimization framework called DRASTIC (Dynamic Resource Allocation for Slicing in Task-aware Closed-loop tactile Internet applications). The proposed framework dynamically allocates resources among network slices supporting both enhanced Mobile Broadband (eMBB) and high reliable low-latency communication (HRLLC) users. The algorithm ensures queue stability and meets delay targets with high probability under a Markov-modulated Poisson traffic, exploiting a Lyapunov guided advantage actor critic reinforcement learning technique. The proposed network model includes an open-loop eMBB queue whose arrival and departure are mainly driven by throughput demand, as well as a closed-loop HRLLC queue that captures feedback and task execution effects. A task execution dependent dexterity index adjusts the effective arrival rate, creating a feedback-aware interaction between the network and the task. A probabilistic delay constraint is incorporated into the objective via Lagrangian relaxation, yielding a min–max optimization framework that enforces latency guarantees while maximizing throughput for both types of users.
Simulation results demonstrate that the proposed framework meets diverse Quality of Service (QoS) requirements, maintains queue stability under dynamic wireless and robotic task variation conditions, and outperforms other approaches.

\end{abstract}

\begin{IEEEkeywords}
6G, Tactile Internet, 
HRLLC, eMBB, Network Slicing, MMPP, probabilistic delay, Lyapunov optimization, reinforcement learning, A2C.
\end{IEEEkeywords}

 \vspace{-2mm}
\section{Introduction}

The evolution of wireless communication has led to the development of 6G technologies, unlocking new paradigms for cyber-physical applications. One of the most transformative applications is the Tactile Internet, which enables real-time, interactive control of remote robots through hyper-reliable low-latency communication (HRLLC) \cite{8329619}, e.g., remote surgery, industrial manipulation, and teleoperated driving. Alongside HRLLC, enhanced mobile broadband (eMBB) services, e.g., video streaming, AR/VR continue to demand high data throughput, creating a complex challenge for radio resource allocation within a shared wireless infrastructure.
In such a setting consisting of coexisting services with conflicting Quality-of-Service (QoS) requirements, network slicing~\cite{habibi2023toward} can be a useful tool for satisfying the demands of different types of applications, along with better and fair spectrum utilization.

\begin{figure*}[!t]
\centering
\includegraphics[width=0.7\linewidth,height=7.5cm]{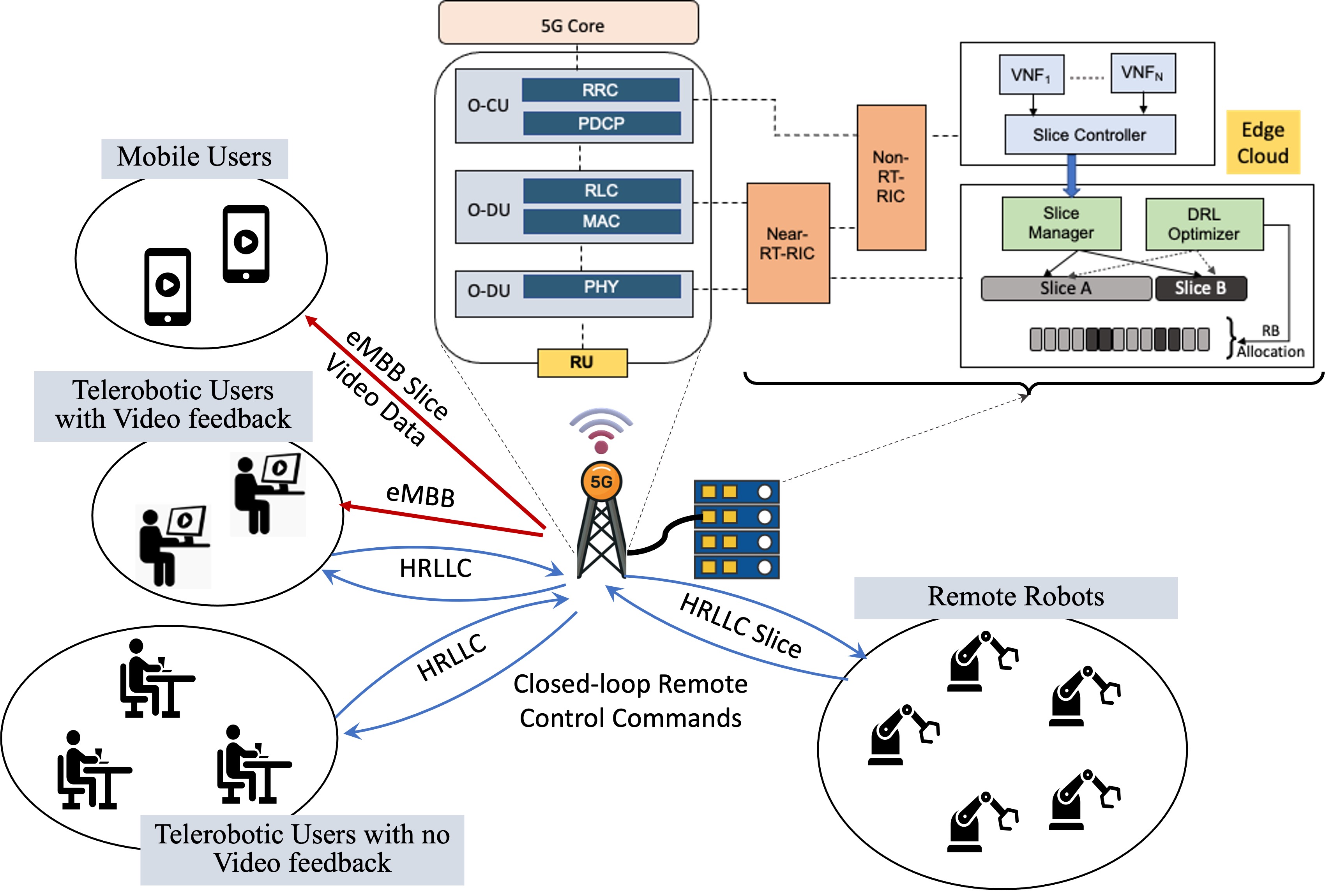}
\caption{6G network slicing for tactile teleoperation: an HRLLC slice carries control/feedback traffic and an eMBB slice carries video traffic, coexisting with background mobile eMBB users under a shared PRB budget}
\label{fig1:topology}
\vspace{-6mm}
\end{figure*}

Fig.~\ref{fig1:topology} shows such a scenario, where
HRLLC slices support haptic and control feedback loops for the remote control of robots, whereas eMBB slices convey high-resolution video streams from the environment back to the users. 
This imposes a critical challenge in ensuring reliable and low-latency control command delivery while maintaining high throughput over limited spectrum, especially under bursty or human-driven application traffic \cite{saad2019vision,lee2020hosting}.
Additionally, tactile Internet applications such as telerobotic traffic exhibit strongly non-Poisson behavior due to human-in-the-loop dynamics.
Such behavior is well captured by Markov-modulated Poisson processes (MMPPs), and it interacts non-trivially with queueing delays and task execution at the robot side. Traditional schedulers such as Round-Robin and Proportional-Fair are not designed to account for MMPP arrivals, probabilistic delay targets, or the tight coupling between communication and control in a closed loop. To bridge this gap, we introduce \textbf{DRASTIC}, a 
deep reinforcement learning (DRL) based dynamic resource allocation framework, integrated with
Lyapunov-based queue stability and task-aware adaptation.

The emergence of Open Radio Access Network (O-RAN) architectures makes this problem even more compelling. O-RAN disaggregates the traditional gNB into virtualized O-CU, O-DU, and O-RU components and introduces RAN Intelligent Controllers (RICs) that expose standardized, programmable control loops at near-real-time (10\,ms-1\,s) and non-real-time ($>$1\,s) time scales \cite{foukas2017orion,arnaz2022toward}. These RICs host AI/ML-powered applications, xApps in the near-RT RIC and rApps in the non-RT RIC, that can dynamically optimize radio resource management, slicing, and QoS policies across heterogeneous services via open interfaces such as E2 and A1 \cite{brik2022deep}. This opens the door to deploying learning-based schedulers for HRLLC/eMBB coexistence as first-class, standard-compliant control applications in the RAN \cite{lotfi2025oran}
by implementing adaptive, O-RAN-compatible schedulers that explicitly target probabilistic delay guarantees under realistic telerobotic traffic.

\subsection{Contributions}

Our contributions through the proposed \textbf{DRASTIC} framework are the following:

\begin{itemize}\itemsep 3pt
\item 
We introduce a task-aware closed-loop HRLLC traffic model by coupling the HRLLC arrival process to a dexterity/complexity index, enabling resource allocation that reflects communication–control interaction. 

\item 
We formulate HRLLC delay reliability as a probabilistic constraint and derive a differentiable exponential surrogate that is embedded in a Lyapunov drift-plus-penalty objective. This yields a min–max optimization framework in which a variable adaptively penalizes delay violations while preserving throughput and queue stability.

\item 
The per-slot physical resource block (PRB) allocation problem is cast as a Markov Decision Process and solved via an Advantage Actor-Critic (A2C) algorithm. The learned policy is designed to operate as an O-RAN compliant xApp on the near-RT RIC, observing queue backlogs, channel conditions, and dexterity indices and issuing scheduling decisions to the O-DU via the E2 interface.

\item 
Extensive simulations show that the proposed Lyapunov–A2C–Lagrangian framework significantly outperforms Round-Robin and Proportional-Fair schedulers in terms of HRLLC delay reliability, even under highly bursty MMPP-driven arrivals.
\end{itemize}

\section{Background and Related Work}

A common limitation in existing 5G/6G slicing and scheduling for URLLC/HRLLC communications is that packet arrivals are treated as exogenous (e.g., independent Poisson or stationary burst models), largely decoupled from the underlying control or task execution process \cite{pang2025communication}. 
{However, in tactile Internet applications, e.g., telerobotics, traffic is inherently \emph{closed-loop}, where the command/feedback generation rate depends on the task being executed and the robot-side responsiveness, which are in turn affected by communication latency and reliability. This creates a feedback interaction in which (i) task dynamics influence the effective HRLLC arrival demand, and (ii) spectrum-resource allocation and delay performance influence subsequent control/feedback behavior. Ignoring this coupling can lead to a systematic mismatch between offered load and allocated service, resulting in unnecessary queue growth and degraded delay reliability during task regime changes.}

In this work, we explicitly capture this communication-control coupling by introducing a task-dependent dexterity index that modulates the HRLLC traffic demand, and by including this task indicator in the scheduler state. As a result, the learned policy can proactively adapt PRB allocation as task complexity changes, matching service to control-induced traffic fluctuations and preserving both queue stability and probabilistic delay reliability.

Complementary research has explored Lyapunov optimization for delay-sensitive and HRLLC traffic. Liu et al.~\cite{liu2018delay} applied Lyapunov drift-minimization to manage delay-constrained wireless networks, while Li et al.~\cite{li2018learning} investigated reinforcement learning for queue-aware performance optimization in edge systems. Although foundational, these approaches consider either HRLLC or eMBB traffic in isolation and do not address robotic workload dynamics. More recent work has begun integrating control-aware metrics into communication scheduling. In~\cite{golmohammadi2024lyapunov}, the authors introduced a dexterity-aware queueing formulation for robotic command execution. In ~\cite{garuda2024closed}, a software-defined experimentation platform is demonstrated for validating telerobotic performance over 5G. Despite these advances, existing approaches remain limited in three key aspects: (i) they do not provide a unified learning framework that jointly enforces communication-layer and control-layer constraints; (ii) they overlook control-influenced arrival and departure processes, including bursty human behavior and task-dependent arrival rates; and (iii) they do not incorporate emerging 6G requirements such as HRLLC reliability, AI-native optimization, and dynamic slice adaptation. 
Consequently, current methods are insufficient for immersive teleoperation, remote surgery, and industrial robot control, where queue instability, variable human actions, and stringent end-to-end delay guarantees must be jointly managed, which our proposed \textbf{DRASTIC} framework offers.

\begin{figure} [t]
\centerline{\includegraphics[width=\linewidth, height = 5.8cm]{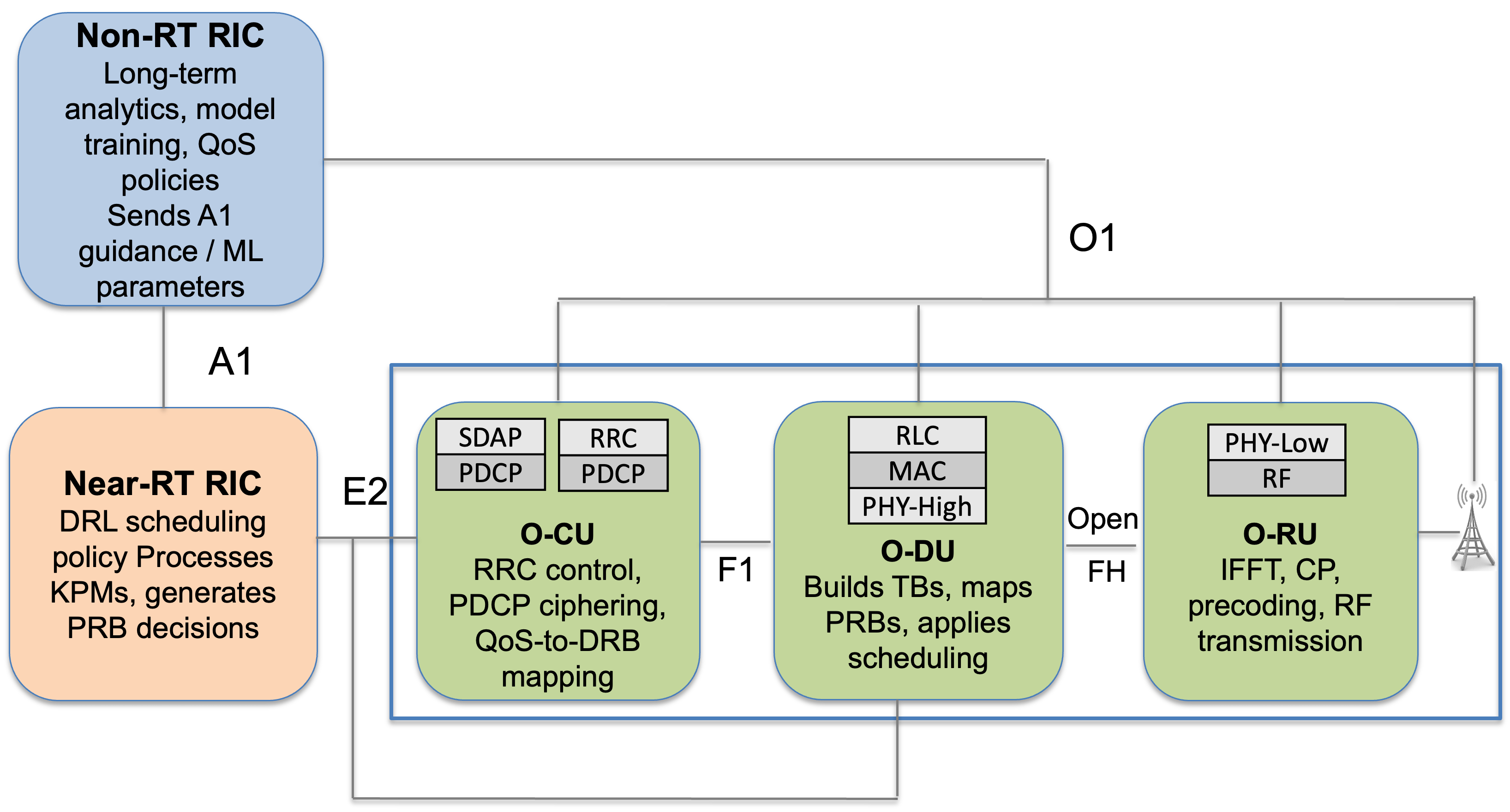}}
\vspace{-2mm}
\caption{O-RAN deployment of DRASTIC}
\label{fig:oran}
\vspace{-4mm}
\end{figure}

Fig.~\ref{fig:oran} illustrates how the proposed scheduler interfaces with the O-RAN control hierarchy and the O-CU/O-DU/O-RU pipeline. The non-RT RIC (timescales$>$1 s) performs long-horizon analytics and model guidance, including slice-level objectives, latency-reliability targets, and policy updates delivered to the near-RT domain via the A1 interface. The near-RT RIC (10-100 ms) hosts DRASTIC as an xApp and executes lightweight inference using live measurements reported by the O-DU, including queue occupancies, PRB utilization, throughput indicators, HARQ feedback, and CQI/KPMs \cite{bonati2020open,bonati2021intelligence}. Based on these real-time observations, the xApp computes per-slot PRB allocation decisions and returns them to the DU for execution. This separation preserves O-RAN timing constraints while enabling closed-loop, slice-aware scheduling that adapts to bursty HRLLC traffic and time-varying channel conditions \cite{polese2023understanding}, thus can support effective resource allocations in applications like tactile Internet.

\vspace{-2mm}
\section{System Description and Problem Formulation}
We consider a 6G system with one gNB serving $U=n_h+n_e$ users, where $n_h$ are HRLLC (telerobotic) users and $n_e$ are eMBB users (Fig.1).  Our objective is to allocate the available bandwidth to both type of user slices. The total bandwidth of $B$ Hz is divided into $K$ PRBs of width $B_k=B/K$. We maintain separate queues for HRLLC and eMBB users. Let $F_i(t)$ denote the HRLLC queue backlog for user $i$ and $G_i(t)$ denote the eMBB queue backlog for user $i$ at the beginning of slot $t$. The following subsections describe the queue dynamics for both set of users.

\subsection {Network Model}
\subsubsection{\textbf{HRLLC Traffic}}
We assume the reference commands for HRLLC end users (agents/robots) are conventionally generated by humans (Fig.~\ref{fig1:topology}). The human generated signals are transferred through the network to the end users may not be uniform (it can be bursty or slow). To capture this behavior in a realistic way, HRLLC arrivals are modeled by a two-state Markov-Modulated Poisson Process (MMPP). Let $S_i(t)\in\{1,2\}$ be the underlying Markov chain with a generator matrix
\vspace{-2mm}
\begin{equation*}
    \mathbf{Q}_i =
\begin{bmatrix}
-\alpha & \alpha \\
\beta   & -\beta
\end{bmatrix},
\end{equation*}

where $\alpha$ and $\beta$ are the transition rates (bursty and slow). Here, $S_i(t)\in\{1,2\}$ denotes the MMPP state of HRLLC user $i$ at slot $t$. 
Given $S_i(t)=s$, the number of arrivals in slot $t$ follows a Poisson distribution with intensity $\lambda_s$ (e.g. $\lambda_2>\lambda_1$). 
Under the transition rates $\alpha$ (from state $1$ to $2$) and $\beta$ (from state $2$ to $1$), the Markov chain has stationary probabilities
\vspace{-2mm}
\[
\pi_1 = \frac{\beta}{\alpha+\beta}, \qquad 
\pi_2 = \frac{\alpha}{\alpha+\beta}.
\]
Therefore, the long-term average arrival rate of user $i$ is the weighted mean
\vspace{-2mm}
\begin{equation*}
    \bar{\lambda}_i=\pi_1\lambda_1+\pi_2\lambda_2
\end{equation*}
\vspace{-2mm}


\textbf{Closing the HRLLC feedback loop:}
To capture the closed-loop nature of telerobotic traffic, we model the dependence of HRLLC packet generation on the robot's/ agent's task execution state. In each slot, the remote robot (or agent) provides a scalar feedback signal, termed the dexterity index $DXI_i$, that captures the task difficulty for user $i$. This feedback is available at the network head (operator side) and is used by the human reference generator to adapt its reference command generation behavior dynamically. Intuitively, when the robot is executing a difficult maneuver (high $DXI_i$), the operator should issue fewer updates and the command arrival should slow down; similarly, when the task is easier (low $DXI_i$), the reference updates can be generated at an ideal rate.

The two-state MMPP modulates its intensity using $DXI_i$. We define:
\vspace{-2mm}
\begin{equation}
\lambda_i(t) = \lambda_{S_i(t)} - \beta\,DXI_i,
\end{equation}
so that a larger $DXI_i$ reduces the instantaneous arrival intensity. The resulting HRLLC arrivals satisfy
\begin{equation}
A_i^h(t) \sim \mathrm{Poisson}\!\left(\lambda_i(t)\right)
= \mathrm{Poisson}\!\left(\lambda_{S_i(t)} - \beta\,DXI_i\right).
\end{equation}

Let $r_i^h(t)$ denote the achieved service rate (departure rate) for the HRLLC user $i$ in time slot $t$, defined in equation~\eqref{eq:datarate}. The HRLLC queue dynamics can be written as
\begin{equation}
F_i(t{+}1)
=
\big[ F_i(t) + A_i^h(t)  - r_i^h(t) \big]^+,
\qquad i \in \mathcal{R},
\label{eq:hrllc_queue_update}
\end{equation}

where $[x]^+=\max\{x,0\}$.

\subsubsection{\textbf{eMBB Traffic}}
We model the eMBB user's data generation (arrival) and service rate (departure) conventionally. Hence, eMBB arrival is modeled as a Poisson distribution with a fixed rate $\lambda_i^e$. Let $A_i^e(t)$ be the arrivals and $r_i^e(t)$ the achieved service rate for the eMBB users obtained by equation~\eqref{eq:datarate}. So, the eMBB queue can be expressed as:
\begin{equation}
G_i(t{+}1)
=
\big[ G_i(t) + A_i^e(t) - r_i^e(t) \big]^+,
\qquad i \in \mathcal{E}.
\label{eq:embb_queue}
\end{equation}
\textbf{Remark 1:} We proposed two queues for two user groups. The eMBB user's queue is an \emph{open-loop} system in the sense that their traffic generation is exogenous. In contrast, HRLLC telerobotic user's queue is a \emph{closed-loop} system since their packet generation depends on the task execution feedback (dexterity index) and therefore can change dynamically over time. In the next section, when we propose a deep reinforcement learning approach to optimal bandwidth allocation, these 2 queues will decide the state transition dynamics for the Markov decision process.

\subsubsection{\textbf{Channel Model and Achievable Rate}}
To model a realistic network behavior, we consider the channel gain to vary across users, PRBs, and time slots due to small-scale fading. Let $h_{ij}(t)$ denote the Rayleigh fading channel gain of user $i$ on PRB $j$ during slot $t$, and let $p_{ij}(t)$ be the transmit power allocated on that PRB. The achievable rate on PRB $j$ for user $i$ is
\vspace{-2mm}
\begin{equation}
r_{ij}(t)=B_k\log_{2}(1 + \frac{p_{ij}(t){h_{ij}}^2(t)}{{\sigma^2}})
\end{equation}
where $\sigma^2$ is the noise variance and $B_k=B/K$ is the PRB bandwidth \cite{mahmud2020performance}. The total rate for user $i$ is
\vspace{-2mm}

\begin{equation}
r_i(t)=\sum_{j=1}^{K}\rho_{ij}(t)r_{ij}(t),
\label{eq:datarate}
\end{equation}
where $\rho_{ij}(t)\in\{0,1\}$ indicates whether PRB $j$ is assigned to user $i$. For consistency with the queue updates, we use $r_i^h(t)=r_i(t)$ for $i\in\mathcal{R}$ (HRLLC) and $r_i^e(t)=r_i(t)$ for $i\in\mathcal{E}$ (eMBB).

\subsection{Problem Formulation}
The gNB has to share a finite bandwidth $B$ (equivalently $K$ PRBs) among two different groups of users while ensuring each user receives the optimal data rate. This resource sharing is challenging because the network varies dynamically over time due to queue and channel condition variation. In addition, the HRLLC slice must satisfy a strict latency reliability requirement, while eMBB users aim for high throughput.

To achieve this objective, at each slot $t$, the scheduler selects an integer PRB allocation vector
\vspace{-2mm}
\begin{equation}
a(t) = \big[a_1(t), a_2(t), \ldots, a_U(t)\big],
\end{equation}
where $U=n_e+n_h$ and $a_i(t)$ denotes the number of PRBs assigned to user $i$ during slot $t$. 

Using the achievable rates induced by the channel model \eqref{eq:datarate}, we aim to optimize the long-term data-rate performance of both HRLLC and eMBB users:
\begin{equation}
{
\arg\max_{a_i}
\lim_{t \to \infty}
\left(
\frac{1}{t} \sum_{\tau=1}^{t}
\sum_{i=1}^{n_h}
\big(r_i^{h}(\tau)\big)^2
+
\frac{1}{t} \sum_{\tau=1}^{t}
\sum_{i=1}^{n_e}
\big(r_i^{e}(\tau)\big)^2
\right)}
\label{eq:original_obj}
\end{equation}
where $r_i^{h}(t)$ and $r_i^{e}(t)$ are the achieved rates for HRLLC and eMBB users, respectively. For numerical stability \eqref{eq:original_obj} can be equivalently reformulated as
\begin{equation}\label{oppf1}
{
{\arg\min_{a_i}}\lim_{t \rightarrow \infty} \left( \sum_{\tau = 1}^{t} \sum_{i=1}^{n_h}  \frac{1}{(r_i^h)^2 (\tau)+\epsilon }+\sum_{\tau = 1}^{t}\sum_{i=1}^{n_e}  \frac{1}{(r_i^e)^2(\tau)+\epsilon}\right)}
\end{equation}
with a small constant $\epsilon>0$.
The optimization is subject to three coupled constraints:
\setlength{\jot}{0pt} 
\begin{subequations}\label{constr1}
\begin{align}
\label{4a}
&Pr (D_{ti} > D_{max}) \leq 1 - \chi_h, \ \forall i = 1, \hdots, n_u, \\ 
\label{4b}&
\sum_{i=1}^{U} a_i(t) = K,\\
\label{4c}
& a_i(t) \in \{1,2,\ldots,K\}.
\end{align}
\end{subequations}
Constraint \eqref{4b} enforces the \emph{instantaneous PRB budget}: all $K$ PRBs must be allocated in every slot. Constraint \eqref{4c} enforces \emph{integer allocations} (each user receives an integer number of PRBs and at least one PRB). Constraint \eqref{4a} enforces the \emph{HRLLC delay-reliability target}: for each HRLLC user, the probability that the end-to-end delay $D_{ti}$ exceeds the deadline $D_{\max}$ must remain below $1-\chi_h$. Because $D_{ti}$ depends on both the queue evolution \eqref{eq:hrllc_queue_update}-\eqref{eq:embb_queue} and the achieved rates \eqref{eq:datarate}, the constraint \eqref{4a} couples the allocation decisions across time slots. As a result, the scheduler must continuously adapt $a(t)$ to changing queues and channels while satisfying \eqref{constr1}.

\noindent
\textbf{Note (HRLLC Probabilistic Delay Constraint):}
For HRLLC user $i$, the total end-to-end delay is
\vspace{-2mm}
\begin{equation}
D_{ti} = D_i^{comm} + D_i^{proc},
\label{eq:delay_definition}
\end{equation}
\vspace{-2mm}

where $D_i^{proc}$ is the processing/propagation delay and $D_i^{comm}$ is the communication delay. The reliability requirement is
\vspace{-2mm}
\begin{equation}
\Pr\!\left(D_i^{comm} > D_{\max} - D_i^{proc}\right) \le 1 - \chi_h,
\label{eq:prob_delay_constraint}
\end{equation}
\vspace{-2mm}
where $D_{\max}$ is the latency budget and $\chi_h$ is the target reliability.

Directly optimizing the tail probability in \eqref{eq:prob_delay_constraint} is difficult because $D_i^{comm}$ depends on time-varying rates and queue backlogs. To obtain a tractable proxy, we define
\vspace{-1mm}
\begin{equation}
y_i(t) = 
\exp\!\Big(\frac{A_ i^h(t)-r_i^h(t)}{p}\big(D_{\max} - D_i^{proc}\big)\Big) 
- (1 - \chi_h),
\label{eq:surrogate_yi}
\end{equation}
where $p$ is a packet size. This surrogate allows incorporating delay reliability together with the queue dynamics \eqref{eq:hrllc_queue_update}-\eqref{eq:embb_queue} and the rate model \eqref{eq:datarate} in an online PRB allocation policy.

\section{Proposed DRASTIC Framework}
DRASTIC couples queue-theoretic stability with learning-based resource allocation for eMBB/HRLLC slicing. At each time slot, the scheduler observes the network state (queues, channel conditions, and task complexity) and outputs an optimal PRB allocation.

\subsection{Lyapunov Drift-Plus-Penalty Reformulation}
\label{subsec:lyapunov_dpp}

A key requirement in this mixed service network is queue stability. The HRLLC and eMBB buffers should not grow without limit, while the scheduler still aims for high efficiency and reliable service. Directly optimizing long-term performance is difficult because traffic and wireless channels change randomly over time, and each scheduling decision affects future queue sizes. To support real-time decisions while keeping stability in mind, we use a Lyapunov drift plus penalty approach.

Let $F_i(t)$ denote the HRLLC backlog of user $i$ and $G_i(t)$ denote the eMBB backlog
of user $i$ at time slot $t$. Collecting all backlogs into the aggregate queue state $Q(t)$,
we define a standard quadratic Lyapunov function
\vspace{-2mm}
\begin{equation}
L(Q(t)) \;=\; \frac{1}{2}\left(\sum_{i}F_i^2(t) \;+\; \sum_{i}G_i^2(t)\right).
\label{eq:lyapunov_function}
\end{equation}
This function can be interpreted as a scalar ``energy'' measure of congestion across the network queue: it increases rapidly when any individual queue becomes large, thereby can help to discourage any
policies that keep the system stable on average by allowing 
persistent backlog accumulation. 

The one-step conditional Lyapunov drift is defined as
\begin{equation}
\Delta L(t) \;\triangleq\;
\mathbb{E}\!\left[\,L(Q(t+1)) - L(Q(t)) \,\middle|\, Q(t)\right].
\label{eq:lyapunov_drift}
\end{equation}
The drift quantifies the expected change in queue ``energy'' from slot $t$ to slot $t+1$,
conditioned on the current backlog state. The conditional expectation is essential because
$Q(t+1)$ depends on stochastic quantities such as random packet arrivals and time-varying channel
conditions. Intuitively, if a scheduling policy keeps $\Delta L(t)$ small, and in particular,
tends to make it negative when queues grow, then the policy prevents growth of the
Lyapunov function over time, which is critical for establishing queue stability.

Minimizing the drift alone enforces stability but does not explicitly encode the desired performance
objective. In addition to bounded backlogs, the scheduler must allocate radio resources to achieve
favorable rate outcomes for both HRLLC and eMBB traffic. To incorporate this system objective into
online decision-making, we introduce a per-slot cost $h_t$ and form the drift-plus-penalty expression
\vspace{-1mm}
\begin{equation}
\Delta L(t) \;+\; V\,\mathbb{E}\!\left[h_t \mid Q(t)\right],
\label{eq:dpp_objective}
\end{equation}
where $V>0$ is a tunable parameter that governs the tradeoff between stabilizing queues and optimizing
performance. Larger values of $V$ place more emphasis on performance at the
expense of potentially higher average queue backlogs, whereas smaller values of $V$ yield more
aggressive backlog reduction and typically smaller queues.

In this work, the instantaneous cost is selected as the per-slot counterpart of the long-term objective function
and is defined as:
\begin{equation}
h_t
=
\sum_{i=1}^{n_h}\frac{1}{\big(r_i^{h}(t)\big)^2+\epsilon}
\;+\;
\sum_{i=1}^{n_e}\frac{1}{\big(r_i^{e}(t)\big)^2+\epsilon},
\label{eq:instantaneous_cost}
\end{equation}
where $r_i^{h}(t)$ and $r_i^{e}(t)$ denote the achieved HRLLC and eMBB rates at slot $t$, respectively,
and $\epsilon>0$ is a small constant introduced for numerical stability. This cost design penalizes the poor
rate outcomes more strongly than moderate ones, encouraging allocations that avoid persistently low rates
service, under unfavorable instantaneous channel conditions.

Finally, to align the drift plus penalty optimization objective with a suitable reinforcement learning (RL) formulation,
We define the per-slot reward as the negative drift plus a penalty quantity:
\vspace{-1mm}
\begin{equation}
R(t) \;=\; -\Big(\Delta L(t) + V h_t\Big).
\label{eq:reward_definition}
\end{equation}
Maximizing the expected cumulative reward, therefore, drives a potential RL learning agent toward actions that
simultaneously (i) reduce the expected growth of the quadratic queue measure, reducing network congestion,
and (ii) improve the throughput (rate-dependent) performance objective captured by $h_t$. Thus, this reward formulation enables an RL-driven online policy to balance reliability-oriented backlog control with throughput-oriented resource
allocation in a unified dynamic optimization framework.

\subsection{Deep Reinforcement Learning Realization}
\label{subsec:drl_realization}
\begin{figure} [!t]
\vspace{-2mm}
\centerline{\includegraphics[width=0.99\linewidth,height=5.5cm]{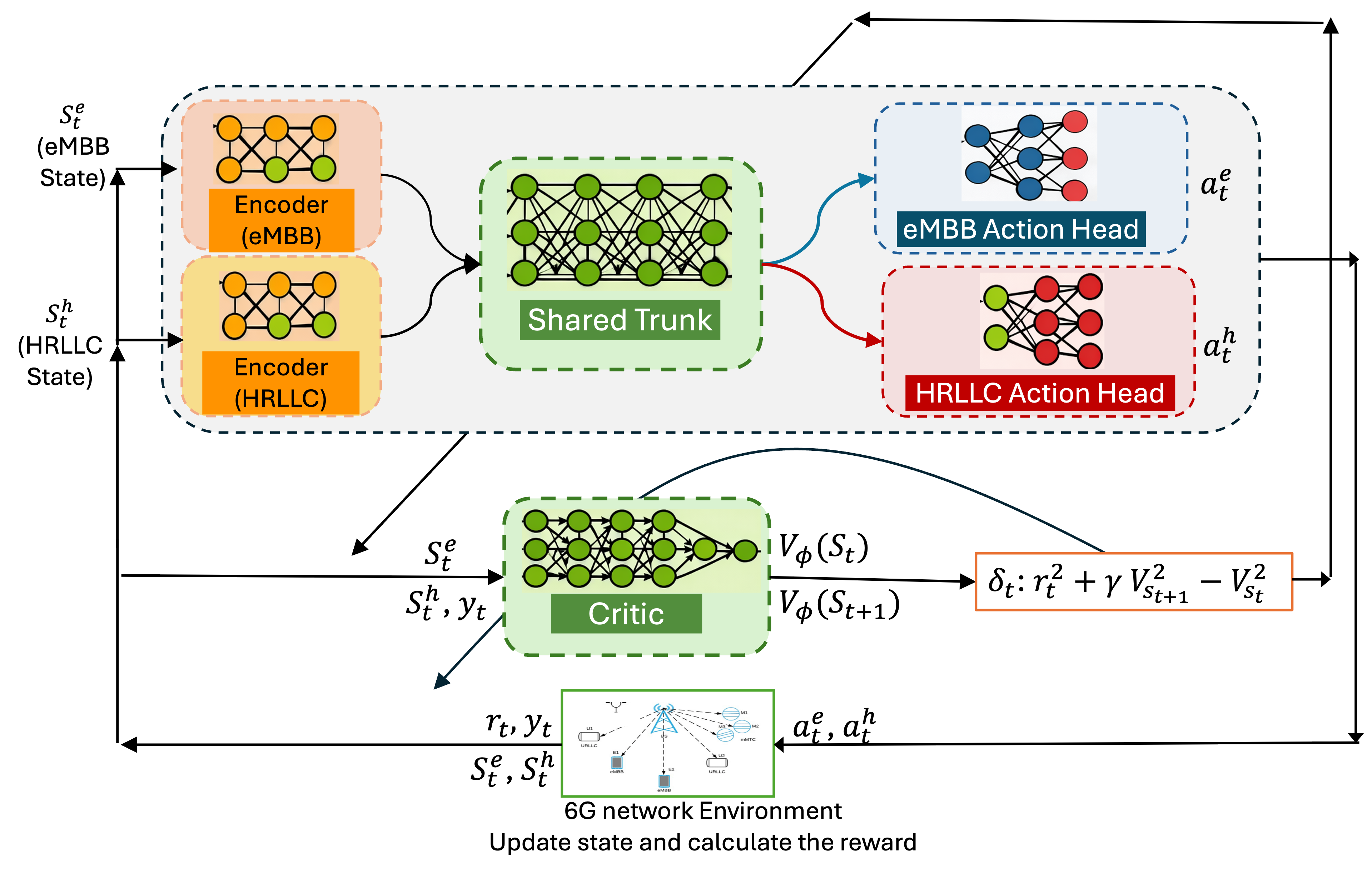}}
\vspace{-2mm}
\caption{DRASTIC architecture and learning loop.}
\label{fig:joint_drl}
\vspace{-6mm}
\end{figure}

To numerically solve the optimization problem defined in \eqref{eq:reward_definition}, we can formulate the
dynamic PRB allocation task as a Markov Decision Process (MDP) and employ a two-head A2C algorithm to learn the mapping from the system state to
resource allocation decisions (actions) \cite{mnih2016asynchronous}. In our implementation as it is shown in Fig.~\ref{fig:joint_drl}, a
\emph{single} joint actor-critic agent is used with a shared representation and two policy heads (one for eMBB and one for HRLLC), along with a single critic value function for the
joint system state.

\subsubsection*{State Space}
At time slot $t$, the agent observes the current network and task status through two
state components: an eMBB state vector $s_t^{e}$ and an HRLLC state vector $s_t^{h}$.

\paragraph{\textbf{eMBB state.}}
The eMBB state is defined as
\begin{equation}
s_t^{e} =
\left[
\mathbf{G}(t),\;
\mathbf{h}^{e}(t),\;
\mathbf{r}^{e}(t),\;
\Delta L^{e}(t)
\right],
\end{equation}
where $\mathbf{G}(t)$ denotes the vector of eMBB queue backlogs, $\mathbf{h}^{e}(t)$
represents the instantaneous eMBB channel conditions, $\mathbf{r}^{e}(t)$ denotes the
achieved eMBB data rates, and $\Delta L^{e}(t)$ is the eMBB drift related term derived
from the Lyapunov framework.

\paragraph{\textbf{HRLLC state.}}
The HRLLC state is defined as
\begin{equation}
s_t^{h} =
\left[
\mathbf{F}(t),\;
\mathbf{h}^{h}(t),\;
\mathbf{r}^{h}(t),\;
\Delta L^{h}(t),\;
y(t),\;
\mathbf{DXI_i}(t)
\right],
\end{equation}
where $\mathbf{F}(t)$ denotes the vector of HRLLC queue backlogs, $\mathbf{h}^{h}(t)$
represents the instantaneous HRLLC channel conditions, $\mathbf{r}^{h}(t)$ denotes the
achieved HRLLC data rates, $\Delta L^{h}(t)$ is the HRLLC drift related term, $y(t)$ is
the instantaneous probabilistic delay violation signal, and
$DXI_i(t)$ is the dexterity-index vector associated with the HRLLC (telerobotic) users,
capturing task complexity.

Finally, the overall state is treated as the pair
\vspace{-1mm}
\begin{equation}
s_t \triangleq \big(s_t^{e},\, s_t^{h}\big),
\end{equation}
which is used by the joint actor-critic network to produce slice-specific allocation
decisions.

\subsubsection{Action Space}

The agent's action at time $t$ is the joint PRB allocation decision, and the space is discrete of the form:
\vspace{-2mm}
\begin{equation}
a_t \triangleq \big(a_t^{e}, a_t^{h}\big),
\end{equation}
where $a_t^{e}$ and $a_t^{h}$ correspond to the eMBB and HRLLC allocation choices,
respectively. These actions determine the achieved rates $\mathbf{r}^{e}(t)$ and
$\mathbf{r}^{h}(t)$, and consequently drive the queue evolution and stability behavior.

\subsubsection{Reward Function}

The reward is designed to reflect the drift plus penalty objective with constraint
enforcement:
\begin{equation}
R(t) =
- \Big( \Delta L(t) + V\, h_t - \lambda_{\ell}(t)\, y(t) \Big),
\label{eq:reward_def}
\end{equation}
where $\Delta L(t)$ is the one-step Lyapunov drift, $h_t$ is the throughput-related
penalty defined in Section~\ref{subsec:lyapunov_dpp}, and $y(t)$ captures the
probabilistic delay-violation signal. The dual variable $\lambda_{\ell}(t)\ge 0$
controls the strength of constraint satisfaction.

\subsubsection{Actor-Critic Updates}
Let $V_{\phi}(s_t)$ denote the critic network parameterized by $\phi$, and let the actor
policy be represented by two categorical heads
$\pi_{\theta}^{e}(a_t^{e}\!\mid s_t)$ and $\pi_{\theta}^{h}(a_t^{h}\!\mid s_t)$ with
parameters $\theta$. Using a one-step bootstrapped target, the TD error is computed as:
\vspace{-2mm}
\begin{equation}
\delta_t = R(t) + \gamma V_{\phi}(s_{t+1}) - V_{\phi}(s_t),
\end{equation}
where $\gamma\in(0,1)$ is the discount factor. The critic is updated by minimizing a
squared (or robust) TD loss, while the actor is updated using the policy gradient based
on $\delta_t$ (advantage estimate). The dual variable is enforced to be nonnegative and
updated online (e.g., via a projected update or a smooth nonnegative parameterization),
ensuring that the learned policy balances queue stability, throughput performance, and
probabilistic delay compliance.

\section{Simulation Results and Performance Analysis}

Simulations were conducted in Python (PyTorch) on an M1 Pro MacBook (16 GB).  
Bandwidth 10 MHz, $K=25$, $N_e=4$, $N_r=3$, $D_{\max}=20$ ms, $\chi_h=0.98$, $\alpha_1=0.1$, $\alpha=0.2$, $\beta=0.2$ s$^{-1}$.  
Discount $\gamma=0.99$, learning rates $10^{-4}$.

\begin{figure} [!t]
\centerline{\includegraphics[width=0.65\linewidth, height=3.4cm]{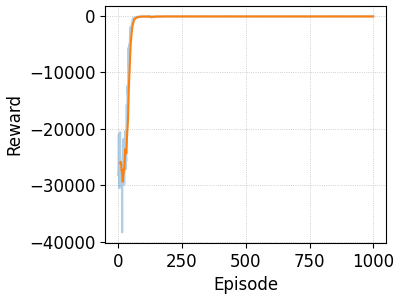}}
\vspace{-2mm}
\caption{Training convergence of DRASTIC (A2C): episodic return versus training episode (with smoothing).}
\label{fig:total_reward}
\vspace{-4mm}
\end{figure}

\begin{figure}[!t]
\centering

\begin{subfigure}[t]{0.49\linewidth}
  \centering
  \includegraphics[width=\linewidth]{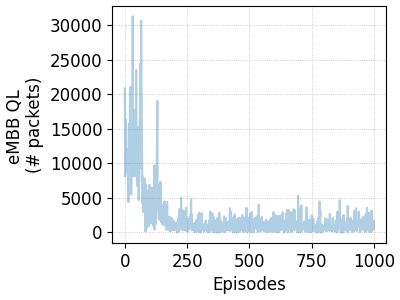}
  \caption{eMBB queue length}
  \label{fig:qle}
\end{subfigure}\hfill
\begin{subfigure}[t]{0.49\linewidth}
  \centering
  \includegraphics[width=\linewidth]{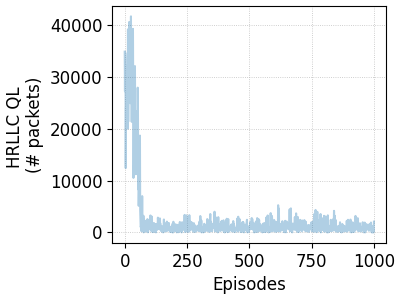}
  \caption{HRLLC queue length}
  \label{fig:qlu}
\end{subfigure}

\caption{Queue evolution during training: eMBB and HRLLC queue backlogs decrease and remain bounded as the learned scheduler converges.}
\label{fig:queues_ab}
\vspace{-4mm}
\end{figure}

\subsection{Convergence Behavior}
\begin{figure} [!t]
\centerline{\includegraphics[width=0.6\linewidth, height=3.5cm]{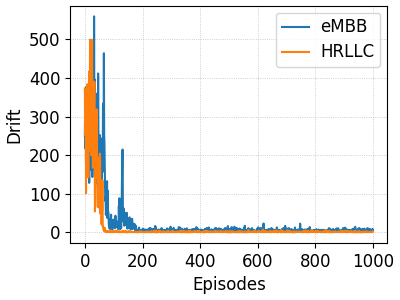}}
\vspace{-2mm}
\caption{Per-slice Lyapunov drift during training: eMBB and HRLLC drift terms decay toward (near) zero, indicating stabilized queue dynamics.}
\label{fig:lya}
\vspace{-6mm}
\end{figure}

During early training, \textbf{DRASTIC} explores widely, which results in inconsistent PRB allocations and occasional delay-reliability violations. This behavior appears as strongly negative episodic returns and large transient growth in both eMBB and HRLLC queue backlogs. As training progresses, \textbf{DRASTIC} learns a more consistent scheduling policy that better matches the stochastic MMPP arrivals with sufficient service, causing both queues to rapidly decrease, Fig.~\ref{fig:queues_ab}(a,b), and remain bounded thereafter, reflected in Fig.~\ref{fig:total_reward}. In parallel, the Lyapunov drift terms, which is shown in Fig.~\ref{fig:lya}, for eMBB and HRLLC decay toward a small steady value, indicating that \textbf{DRASTIC} improves stability (not only reward). After approximately 125 episodes, the return curve plateaus and queue lengths fluctuate within a narrow range, suggesting that \textbf{DRASTIC} has converged to a stable operating regime with only minor residual variations due to channel randomness and bursty arrivals.

\begin{figure} [!t]
\centerline{\includegraphics[width=0.6\linewidth, height=3.5cm]{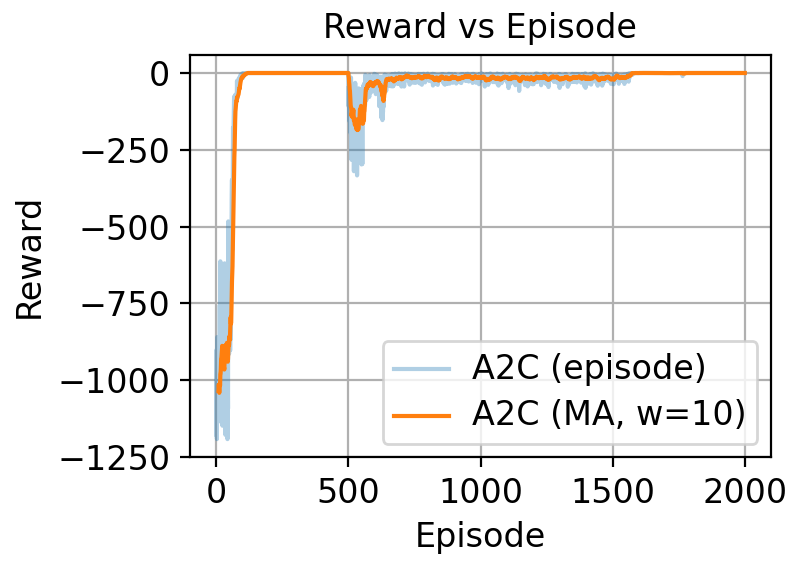}}
\vspace{-2mm}
\caption{Reward under the two-step dexterity experiment.}
\label{fig:two_step_reward}
\vspace{-4mm}
\end{figure}





\subsection{Two Step Dexterity  Index}
\begin{figure}[!t]
\centering

\begin{subfigure}[t]{0.99\linewidth}
  \centering
  \includegraphics[width=\linewidth, height=2.8cm]{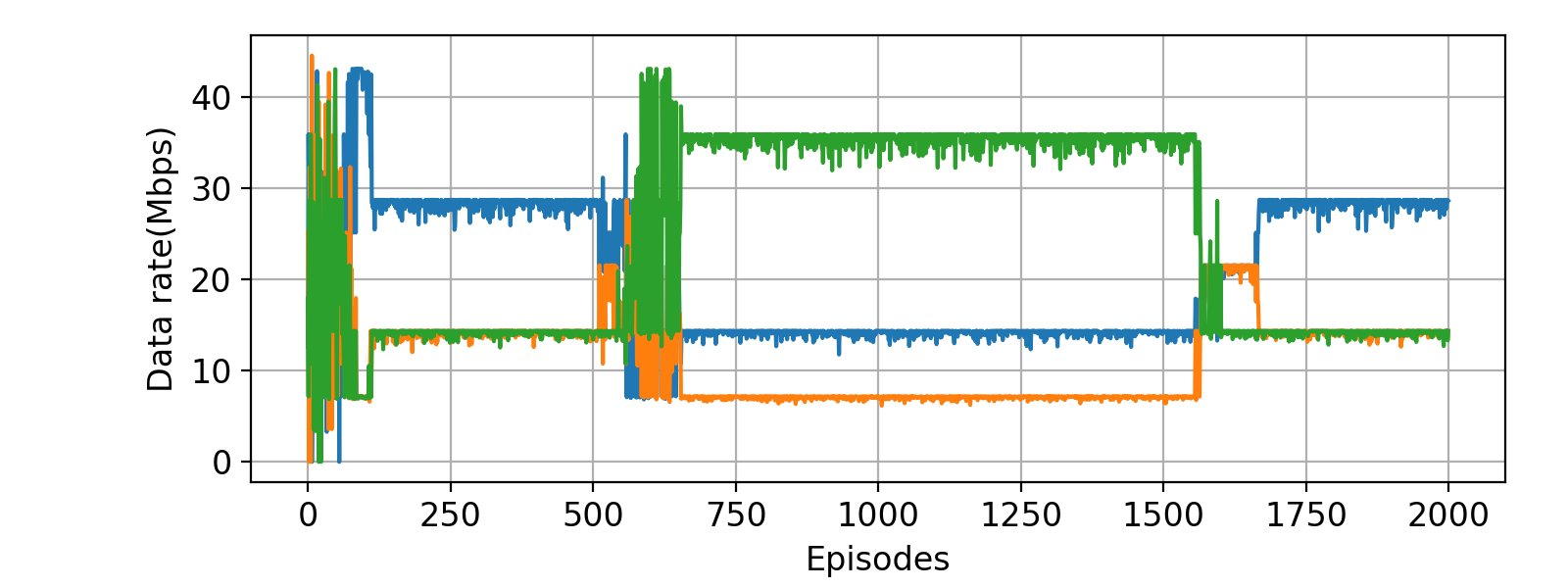}
  \caption{}
  \label{fig:two_step_hrllc_datarate}
\end{subfigure}\hfill

\begin{subfigure}[t]{0.99\linewidth}
  \centering
\includegraphics[width=\linewidth,height=2.8cm]{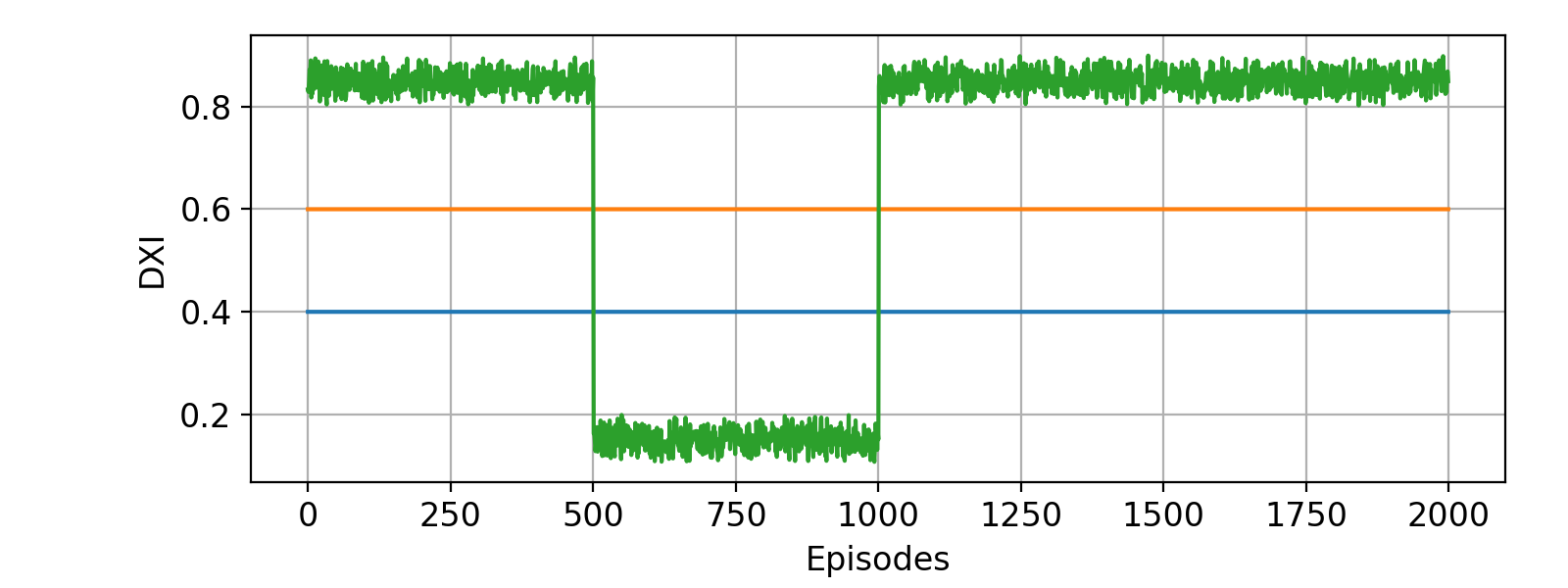}
  \caption{}
  \label{fig:two_step_dex}
\end{subfigure}

\vspace{-2mm}
\caption{Two-step dexterity experiment: (a) HRLLC achieved data rate; (b) imposed two-step dexterity profile. DRASTIC reallocates PRBs to track the demand shift while preserving queue stability and delay reliability.}
\label{fig:two_step_results}
\vspace{-4mm}
\end{figure}

The time-domain results illustrate task-dependent adaptation of the learned scheduler under heterogeneous traffic. While eMBB traffic is open-loop and throughput-driven, HRLLC traffic is closed-loop and coupled to control/feedback via the dexterity index, which modulates the command/feedback generation rate. When the dexterity index decreases, Fig.~\ref{fig:two_step_results}(b), (simpler task), the HRLLC arrival rate increases; the scheduler reallocates more PRBs to the affected user, boosting its data rate and departures to prevent sustained queue growth. The temporary rise in the Lyapunov-related signal around the step change reflects the short-term mismatch between the prior allocation and the new traffic regime, degradation in reward in Fig.~\ref{fig:two_step_reward} at episode 500, and it decays once resources are adjusted and backlog pressure is relieved.

When the dexterity index increases again (more complex tasks), the feasible closed-loop command frequency drops, and HRLLC arrivals decrease. The policy correspondingly reduces that user’s PRB share and redistributes resources while preserving HRLLC latency-reliability targets reflected in Fig.~\ref{fig:two_step_results}(a). Overall, the step responses in arrivals, rates, departures, and drift confirm the intended feedback queue allocation coupling: task complexity reshapes closed-loop demand, and DRASTIC tracks these regime shifts through adaptive bandwidth allocation.


\vspace{-2mm}
\subsection{DRL comparison}
We evaluate three DRL-based schedulers for joint PRB allocation across eMBB and HRLLC traffic. At each time step, the scheduler observes the network state (including traffic and queue-related features) and selects a PRB allocation that partitions a fixed PRB budget among users in both slices while satisfying per-user minimum allocation constraints. The objective is to maximize long-term performance while discouraging delay/queue violations through a penalty term weighted by a Lagrange multiplier $\lambda$.

The compared methods include \textbf{A2C} (discrete policy), \textbf{DQN} (discrete value-based), and \textbf{DDPG} (continuous actor-critic). A2C samples from a categorical distribution over feasible integer allocations, DQN selects an allocation by maximizing estimated action values, and DDPG outputs continuous resource-splitting proportions (separate outputs per slice) that are projected to integer PRB allocations before execution.

\begin{figure} [!t]
\centerline{\includegraphics[width=0.7\linewidth,height=3.4cm]{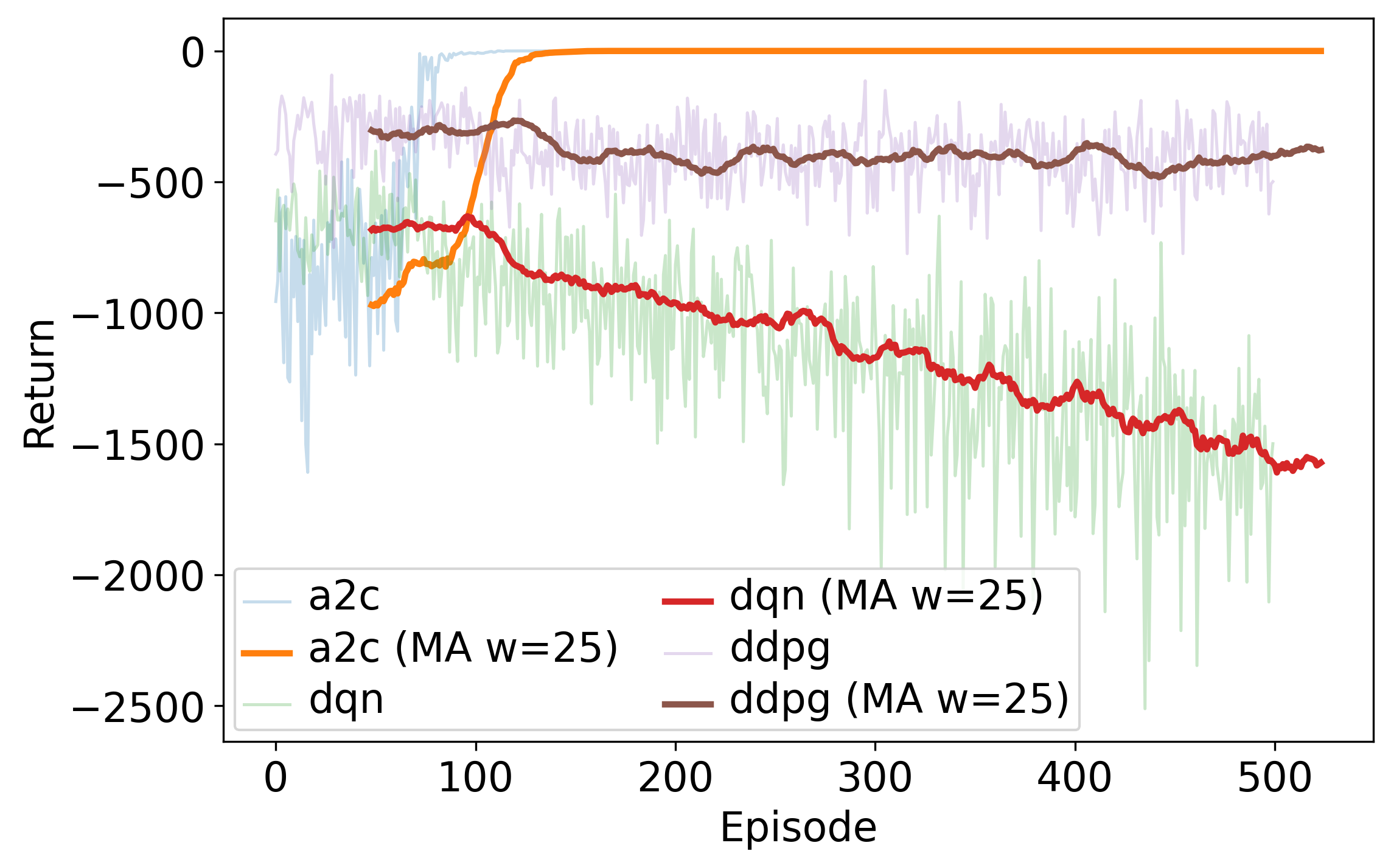}}
\caption{DRL scheduler comparison under identical PRB and channel settings: episodic return for A2C (DRASTIC), DQN, and DDPG, highlighting more stable learning for A2C under discrete PRB allocation constraints.}
\label{fig:drl}
\vspace{-4mm}
\end{figure}

As shown in Fig~\ref{fig:drl}, A2C exhibits stable learning in this setting, whereas DQN and DDPG show limited or no convergence under the same dynamics and constraints. This is largely driven by the discrete nature of the executed allocation. For DQN, the feasible allocation space grows rapidly with the PRB budget and number of users, making exploration and value estimation difficult, particularly when constraint-violation penalties dominate the return. For DDPG, the continuous actor output must be mapped to integer allocations through a non-smooth projection, where many continuous actions map to the same executed allocation, weakening the learning signal and destabilizing critic-actor updates when penalties are large.

\begin{figure}[!t]
\centering

\begin{subfigure}[t]{0.5\linewidth}
  \centering
  \includegraphics[width=\linewidth,height=3.5cm]{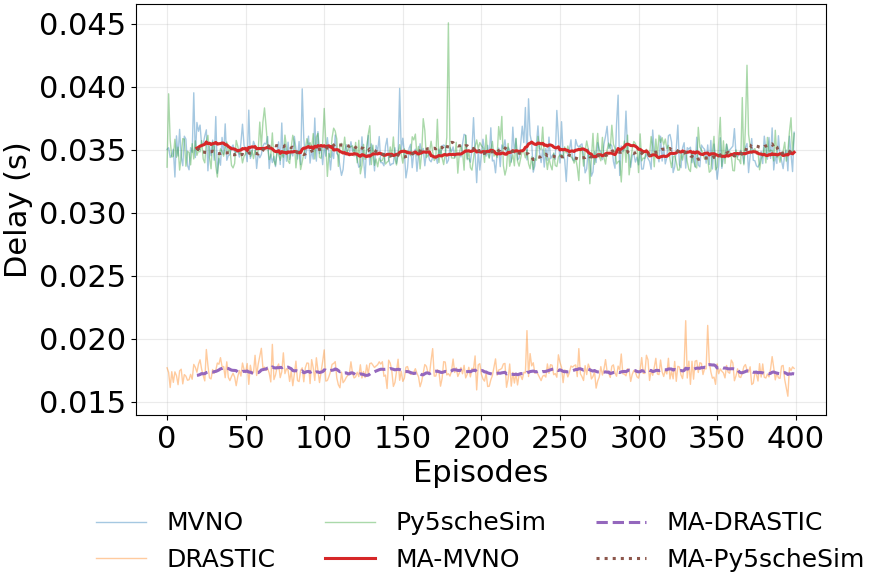}
  \caption{}
  \label{fig:RA}
\end{subfigure}\hfill
\begin{subfigure}[t]{0.5\linewidth}
  \centering
  \includegraphics[width=\linewidth,height=3.5cm]{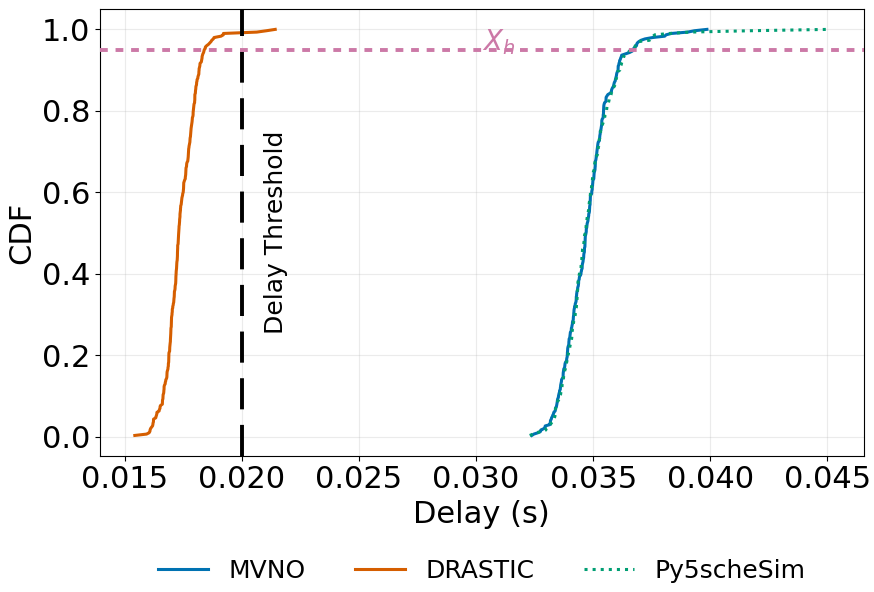}
  \vspace{-2mm}
  \caption{}
  \label{fig:cdf_comparison}
\end{subfigure}

 \vspace{-4mm}
\caption{Delay comparison of DRASTIC against intra-slice baselines MVNO-PF and Py5cheSim-RR. 
(a) Per-episode delay with moving-average (MA) smoothing; DRASTIC achieves less delays compared to PF/RR
(b) CDF of per-episode delay for different allocation strategies.}
\label{fig:RA_and_CDF}
\vspace{-6mm}
\end{figure}

\subsection{Delay and Reliability}
Fig.~\ref{fig:RA_and_CDF} evaluates delay performance and delay-reliability compliance. 
The \emph{per-episode delay} in Fig.~\ref{fig:RA_and_CDF}(a) is computed from the queueing model and reflects the experienced latency under the selected PRB allocation. 
Fig.~\ref{fig:RA_and_CDF}(b) shows the empirical CDF of the communication delay for \textbf{DRASTIC} and the two intra-slice baselines (\textbf{MVNO} and \textbf{Py5cheSim}). The vertical and horizontal dashed lines indicate the communication delay threshold (here $D=0.02$~s) and the reliability target $X_h$ (e.g., $X_h=0.98$), respectively, visualizing the requirement $\Pr(D^{comm}<D_{\max} - D^{proc}) \ge X_h$. DRASTIC meets the $98\%$ reliability target at (or before) $D_{\text{th}}$, while MVNO and Py5cheSim reach the same reliability level only at larger delays, implying a higher probability of threshold violation. All methods are evaluated under identical conditions (same channel model, PRB budget, dexterity-driven HRLLC arrivals, and queue dynamics), so differences reflect only the scheduling policy. We compare \textbf{DRASTIC} against two \emph{intra-slice} baselines: \textbf{MVNO-PF}, which applies a proportional-fair priority rule, and \textbf{Py5cheSim-RR}, which performs channel-agnostic round-robin sharing; neither baseline explicitly optimizes delay reliability.
\cite{pereyra2022open,cannata2024towards}.

\subsection{{Task-complexity (dexterity) sensitivity study:}}
To evaluate whether the proposed formulation captures task-dependent closed-loop traffic, we consider a scenario with five HRLLC users assigned distinct dexterity indices, ranging from low to high task complexity. 

\begin{figure} [!t]
\centerline{\includegraphics[width=0.75\linewidth,height=3.5cm]{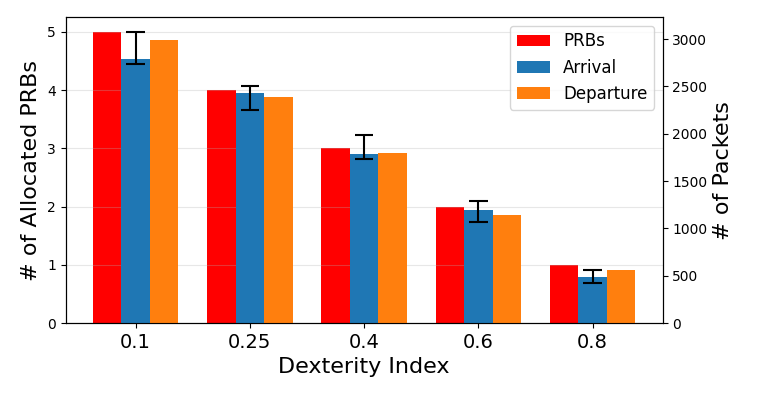}}
\caption{Adaptive bandwidth allocation under task-dependent traffic: PRBs track arrival/departure rates across different dexterity levels.}
\label{fig:dex_invest}
\vspace{-6mm}
\end{figure}

Fig.~\ref{fig:dex_invest} summarizes the resulting steady-state behavior. As the dexterity index increases, the measured arrival rate decreases, and the learned policy correspondingly assigns fewer PRBs to that user. This reduction in allocated bandwidth leads to a lower service capability, which is reflected by the reduced departure rate. Conversely, users with smaller dexterity indices exhibit higher arrivals and are granted a larger PRB share, yielding higher departures to prevent queue build-up. The close tracking between arrivals and departures across dexterity levels indicates that the proposed scheduler consistently matches service rates to task-induced traffic demand, thereby demonstrating sensitivity to task complexity and effective resource adaptation under heterogeneous closed-loop HRLLC workloads.

\section{Conclusion and Future Work}
This paper presented a Lyapunov-A2C-Lagrangian framework for 6G telerobotic scheduling with MMPP-driven arrivals and probabilistic delay guarantees.  
By uniting stochastic optimization and reinforcement learning, the method stabilizes queues, enforces reliability, and maximizes throughput simultaneously.  
The adaptive dual variable acts as an online reliability controller, while the A2C agent approximates optimal drift minimization.

Key outcomes include: (1) stable queue dynamics under correlated bursts; (2) 98\% delay reliability; and (3) throughput gains over conventional schedulers.  
Future extensions include distributed multi-cell coordination, hierarchical slice management including mMTC, and real SDR-based testing using EdgeRIC platforms.  
This framework lays the foundation for autonomous and intelligent RAN scheduling under 6G and beyond.


\bibliographystyle{IEEEtran}
\bibliography{ref}

\end{document}